# Peculiarities of the Solar-Wind/Magnetosphere Coupling in the Era of Solar Grand Minimum

*Yuri I. Yermolaev, Irina G. Lodkina, Aleksander A. Khokhlachev, Mikhail Yu. Yermolaev*

*Space Research Institute, Moscow, Russia*

**Abstract**
Based on the data of the solar wind (SW) measurements of the OMNI base for the period 1976-2019, the behavior of SW types as well as plasma and interplanetary magnetic field (IMF) parameters for 21-24 solar cycles (SCs) is studied. It is shown that with the beginning of the Era of Solar Grand Minimum (SC 23), the proportion of magnetic storms initiated by CIR increased. In addition, a change in the nature of SW interaction with the magnetosphere could occur due to a decrease in the density, temperature, and IMF of the solar wind.

**INTRODUCTION**

The state of the magnetosphere is determined by two competing processes: excitation by external, interplanetary, drivers and relaxation due to internal, magnetospheric, processes. The following disturbed phenomena of the solar wind (SW) are considered as the main interplanetary drivers, which may contain the southward component of the interplanetary magnetic field (IMF) and be geoeffective: interplanetary coronal mass ejections (ICMEs, in this paper we consider two subtypes of ICMEs, magnetic clouds and ejecta, together), compression areas before ICME, sheaths, and compression area between fast and slow SW streams, corotating intertplanetary regions (CIRs) (Gonzalez et al., 1999; Yermolaev et al., 2005; 2021a; Borovsky & Denton 2006).

Solar scientists have long drawn attention to a noticeable decrease in solar activity, the so-called Solar Grand Minimum (Feynman & Ruzmaikin 2011; Zolotov & Ponyavin 2014), but solar wind and magnetosphere specialists relatively rarely pay attention to this fact (for example, Oh & Kim, 2013; Gopalswamy et al., 2015) and usually do not take into account the possible change in the properties of SW and interplanetary drivers during the space era. Recently, on the basis of data from the OMNI database for solar cycles (SCs) 21–24, we showed (Yermolaev et al., 2021b) that in the interval between solar cycles 22 and 23, the structure of the heliosphere changed, and most of the SW and IMF parameters fell by 20–40%, and this drop was observed in all types of SW and all phases of SCs. These changes in SW could not but lead to changes in Solar-Wind/Magnetosphere Coupling, in particular, to a sharp decrease in the number of magnetic storms in SCs 23 and 24. In this short paper, we consider some experimental data on the SW and the magnetosphere related to the weakening of the SW and its effect on solar-terrestrial links.

**DATA AND METHODS**

In this work, we use two sources of information for 1976-2019: (1) Hourly data of solar wind parameters in the OMNI base (https://spdf.gsfc.nasa.gov/pub/data/omni/low_res_omni, King and Papitashvili, 2005), and (2) Intervals of different types of SW in the catalog of large-scale phenomena (http://www.iki.rssi.ru/pub/omni, Yermolaev et al., 2009), created on the basis of the OMNI database.

Due to the relatively small number of magnetic clouds, we analyzed them together with ejecta as a general class of ICME drivers, and did not separate sheath compression areas before MC and ejecta. The parameters of the solar wind are highly variable, and when calculating the average parameters, rather large standard deviations were obtained. However, given that a large number of measurement points (up to several thousand) were used, statistical errors (standard deviations divided by the square root of the number of points) were small, and the obtained average values of the parameters have a high degree of statistical significance (Yermolaev et al., 2021b). In particular, the statistical errors do not exceed the size of the symbols in the figures below.

When associating interplanetary and magnetospheric phenomena, it was assumed that the magnetic storm was generated by the type of interplanetary driver, inside which the minimum of Dst index fell.

**RESULTS AND DISCUSSION**

Recently we showed (Yermolaev et al., 2021b) that in SC 23 and 24, the number of ICMEs and sheaths dropped significantly, while the number and distribution of CIRs did not change significantly. This is confirmed by the quantitative data of the bottom 2 panels of Figure 1, which show the fraction of time from the total observation time during which different disturbed SW types are observed: CIR (left panel) and ICME and sheath (right panel), respectively. The CIR time share increased slightly from ~6 to ~8%, while the ICME and sheath time shares fell from ~18 to ~8% and ~4 to ~2%, respectively. The total number of disturbed SW types fell from 28 to 18%. An interesting feature is observed: if the ICME and sheath time fractions have a common global decreasing trend over SC 21–24 period, then they increase in the SC 22–23 interval, and this interval coincides with the beginning of the fall in the SW and IMF parameters (Yermolaev et al., 2021b ). The data from the top 2 panels of Figure 1 show the proportion of medium and strong magnetic storms (Dst < -50 nT) generated respectively by CIR (left panel) and ICME and sheath (right panel) events. These panels show that against the background of a general drop in the number of magnetic storms (Yermolaev et al., 2021b), the proportion of storms generated by ICMEs and sheaths remains almost unchanged (~35% for ICME and ~15% for sheath events), while the proportion of CIR-induced storms markedly increases from ~17 to ~30%. These dependencies also have features in the SC 22–23 interval.

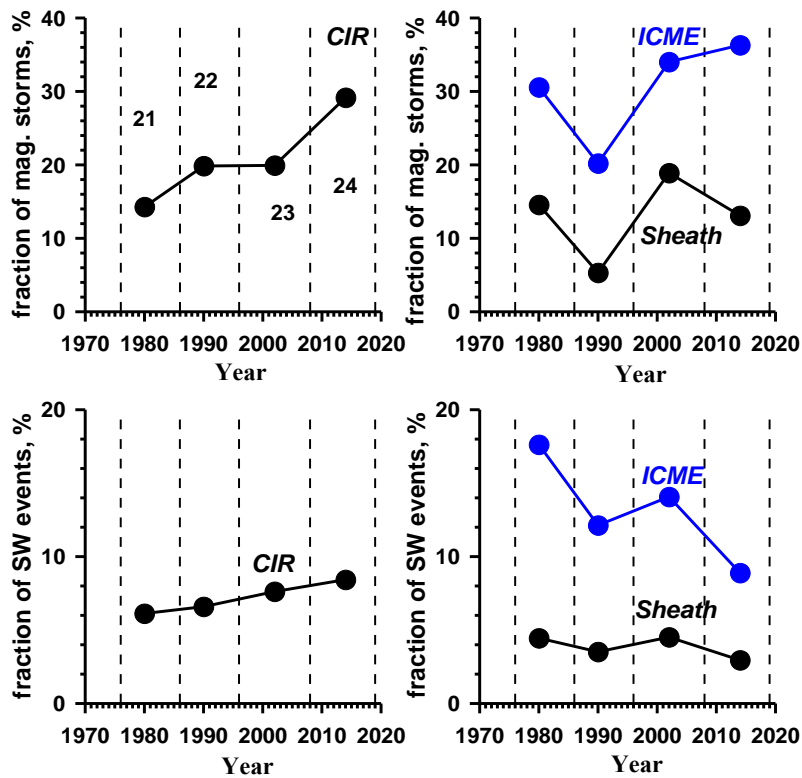

Figure 1. Temporal profiles of fraction of magnetic storms (upper panels) and fraction of SW events (bottom panels) during solar cycles 21-24: for CIRs (left panels) and sheaths (right panels, black circles) and ICMEs (right panels, blue circles)

The fact noted above that the number of storms does not change easily in proportion to the number of corresponding SW drivers can be associated with changes in the values of the SW and IMF during SC 21–24. Figure 2 presents the average values of the density, velocity, IMF magnitude, and solar wind proton temperature for 4 SCs. Only the speed remains approximately unchanged, the other 3 parameters are significantly reduced. How these (and some other) parameters decrease at certain phases of solar cycles and in certain types of SW, including the main interplanetary drivers of magnetospheric disturbances, can be seen in the corresponding figures in paper (Yermolaev et al., 2021b). Here we want to note only some important general facts.

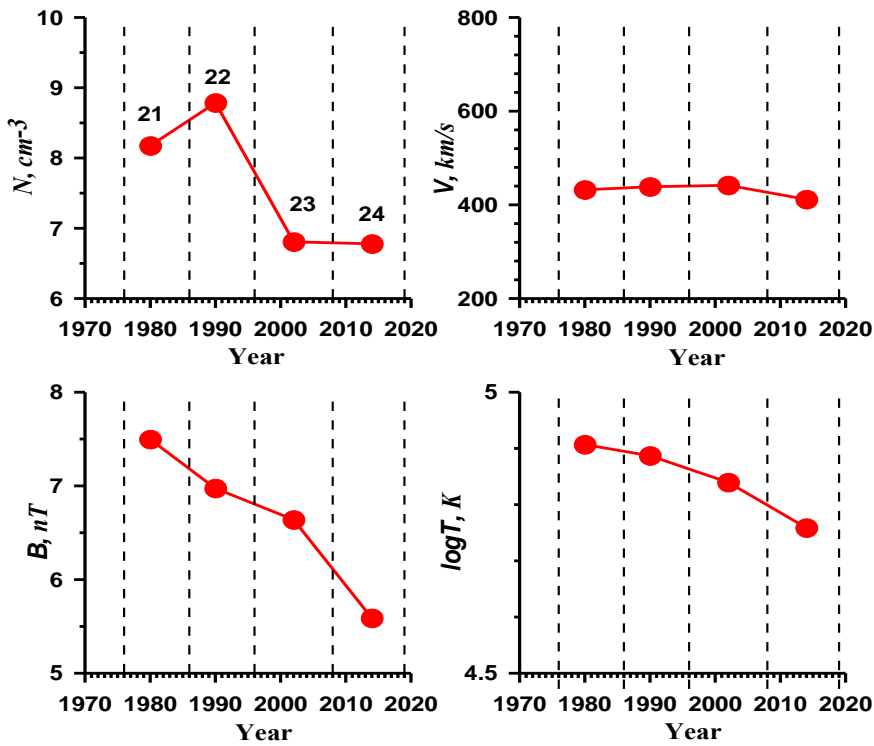

Figure 2. Temporal profiles of density N, velocity V, magnitude IMF B and proton temperature T during solar cycles 21-24

(1) Since in SC 23-24 the temperature and IMF value dropped at a constant velocity, the sonic and Alfvén Mach numbers in the solar wind increased, i.e. at present epoch, the magnetosphere is flowed around by more "supersonic" and more "super-alfvenic" SW than in the previous 21-22 SCs. In addition, the drop in the beta parameter (not shown here, but presented in paper (Yermolaev et al., 2021b)) indicates that the drop in magnetic pressure turned out to be greater than the drop in thermal pressure. Therefore, the regime of the SW flow around the magnetosphere of the Earth and their interaction may be different than in the previous period.

(2) Since the density and dynamic pressure dropped in SC 23-24 by ~30%, the pressure balance of the interplanetary plasma and the magnetospheric plasma at the magnetopause began to be located at larger distances from the center of the Earth, i.e. the size of the magnetosphere and its structures has increased, and the pressure inside the magnetosphere has fallen. This can affect both the nature of SW interaction with the magnetosphere and the physical processes inside the magnetosphere (Antonova et al., 2014; Gopalswamy et al., 2015; Kirpichev et al., 2017).

**CONCLUSIONS**

Thus, an analysis of measurements of SW parameters during 1976-2019 showed that the fall in solar activity at SC 23-24 (in the Era of Solar Grand Minimum) is accompanied by a weakening of the characteristics and a change in the structure of the interplanetary medium. This, in particular, is accompanied by an increase in the number of events in the interplanetary medium and the relative contribution of CIR compression regions to the excitation of magnetic storms against the background of a decrease in the number of ICME and sheath events. In addition, the SW density, temperature, and IMF decrease significantly, which can lead to a change in the nature of the solar wind flow around the magnetosphere and their interaction. Therefore, the data of direct measurements on Solar-Wind/Magnetosphere Coupling obtained during the space era may differ for physical reasons, and their analysis must be divided into the initial period of the space era and the period of the Solar Grand Minimum. Accounting for changes in numbers and properties of external drivers is also important for predicting space weather effects.

**FUNDING**

The work was supported by the Russian Science Foundation, grant 22-12-00227.

**ACKNOWLEDGMENTS**

Authors thank creators of databases https://spdf.gsfc.nasa.gov/pub/data/omni/low_res_omni and http://www.iki.rssi.ru/pub/omni for possibility to use in the work.